\documentclass[pdflatex,sn-aps]{sn-jnl}
\usepackage{graphicx} 
\usepackage{multirow}%
\usepackage{amsmath,amssymb,amsfonts}%
\usepackage{amsthm}%
\usepackage{mathrsfs}%
\usepackage[title]{appendix}%
\usepackage{xcolor}%
\usepackage{textcomp}%
\usepackage{manyfoot}%
\usepackage{booktabs}%
\usepackage{algorithm}%
\usepackage{algorithmicx}%
\usepackage{algpseudocode}%
\usepackage{listings}\usepackage{rotating} 
  \usepackage{tcolorbox}
\usepackage[
singlelinecheck=false 
]{caption}

\usepackage{etoolbox}   
\makeatletter
\patchcmd{\@maketitle}{\artauthors}{\centering{\artauthors}}{}{}
\makeatother

\begin{document}
\title{\centering Comment on \\ ``Sustainability Strategy for the Cool Copper Collider''}
\subtitle{\small (arXiv:2307.04084 [hep-ex], PRX Energy 2 (2023) 4, 047001,  M. Breidenbach et. al.)}

\author[1]{\fnm{C.} \sur{Grojean}}
\author*[2]{\fnm{P.} \sur{Janot}}\email{patrick.janot@cern.ch}

\affil[1]{\small Deutsches Elektronen-Synchrotron DESY, Notkestr. 85, 22607 Hamburg, Germany,\\
Institut für Physik, Humboldt-Universit\"at zu Berlin, 12489 Berlin, Germany}

\affil[2]{\small \orgname{CERN}, \orgdiv{EP Department}, \orgaddress{\street{1 Esplanade des Particules}, \city{Meyrin}, \country{Switzerland}}}

\abstract 
{The paper entitled ``Sustainability Strategy for the Cool Copper Collider'' by M. Breidenbach et al. defines a metric to weigh the electricity consumption and the carbon footprint of future Higgs factory concepts. We show that this metric is flawed in many respects and gives an incorrect representation of reality. We also demonstrates that, irrespective of the drawbacks of this particular estimator, the strategy consisting in using a metric as a multiplicative coefficient of the electricity consumption or the carbon footprint of an ensemble of colliders is fragile at best, and valueless when it comes to arguing in favour of or against such or such future Higgs Factory concept.}

\maketitle

\vfill\eject

\begin{figure}[h]
\centering
\includegraphics[width=0.99\textwidth]{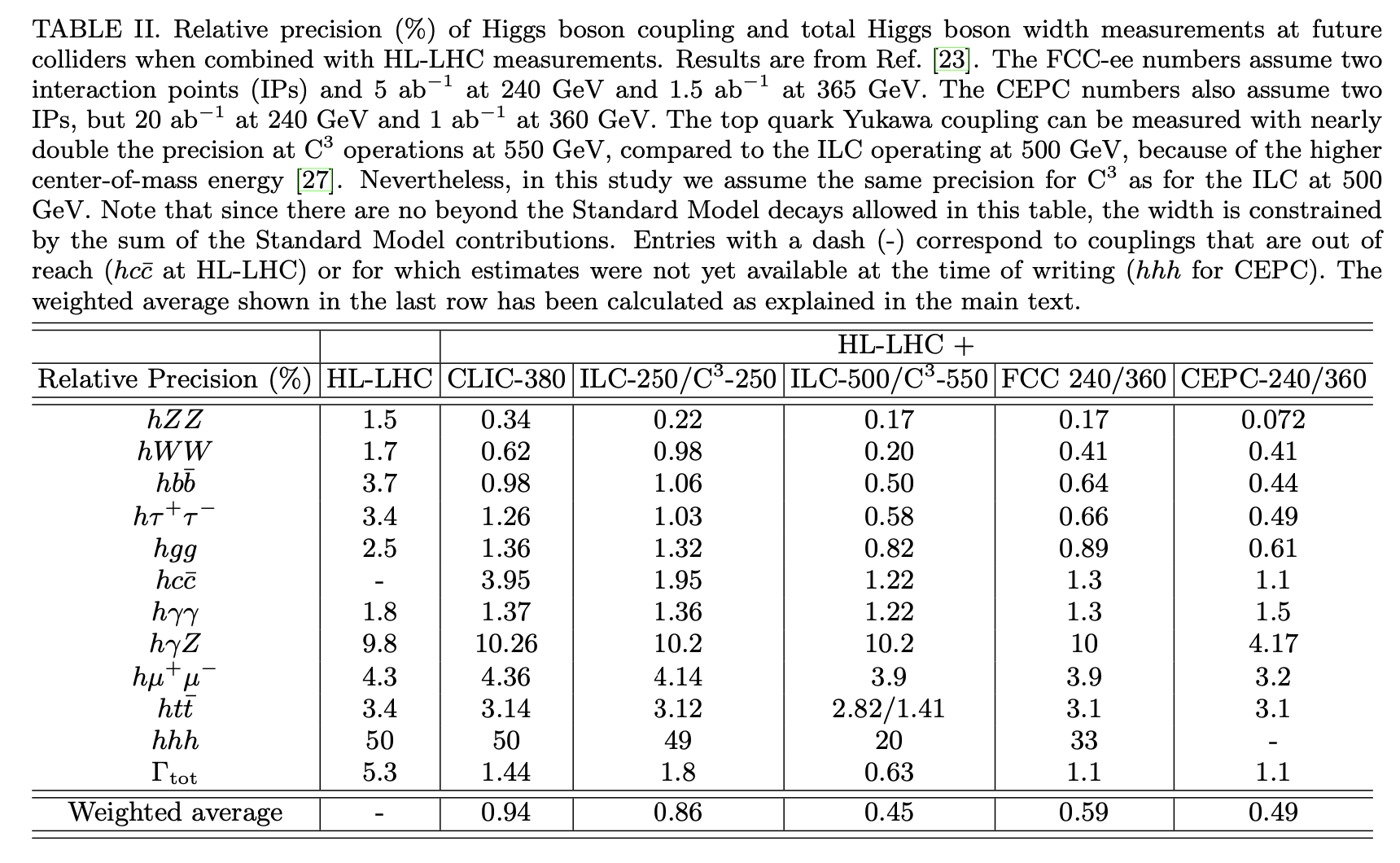}
\caption{\label{fig:screenshot} \small Screenshot of the table with individual Higgs coupling precisions used in Ref.~\cite{Breidenbach:2023nxd}. This table is actually copied from Ref.~\cite{Dawson:2022zbb}, labelled [23] in the original table caption, except for the last and the ante-penultimate rows.
}
\end{figure}

In Ref.~\cite{Breidenbach:2023nxd}, a metric is proposed for the comparison of the electricity consumption and the carbon footprint ``per unit of physics output'' for several Higgs factory (HF) concepts.
According to the authors, this is achieved {\it by taking the average of the relative precision over all Higgs couplings, weighing them by the relative improvement in their measurement with respect to HL-LHC}. The weight of each coupling is defined as 
\begin{equation}
    w_i = \frac{\left( \frac{\delta\kappa}{\kappa} \right)_{\rm HL-LHC, i} - \left( \frac{\delta\kappa}{\kappa} \right)_{\rm HL-LHC+HF, i}}{\left( \frac{\delta\kappa}{\kappa} \right)_{\rm HL-LHC+HF, i}},
    \label{eq:weight}
\end{equation}
and the averaged coupling precision reads
\begin{equation}
    \langle\frac{\delta\kappa}{\kappa}\rangle = \frac{\sum_i w_i \left( \frac{\delta\kappa}{\kappa} \right)_{\rm HL-LHC+HF, i} }{\sum_i w_i}, 
    \label{eq:weighted}
\end{equation}
where $\kappa_i$ is the multiplicative modifier of coupling $i$ and $\delta\kappa_i$ is the precision on this modifier as obtained from a so-called ``$\kappa$ fit''~\cite{deBlas:2019rxi}. The ``raw'' carbon footprint of each Higgs factory is then multiplied by this average coupling precision, chosen as a metric to rank the colliders, in the hope that the outcome of this multiplication is indeed ``a carbon footprint per unit of physics output''. For completeness, a screenshot of the table with the individual coupling precision from Ref.~\cite{Breidenbach:2023nxd} is shown in Fig.~\ref{fig:screenshot}.

We identified a number of serious issues with the table of precisions, the metric, and the weighted average of the coupling precisions, a non-exhaustive list of which follows. 
\begin{enumerate}
    \item First of all, we cannot reproduce the weighted average values shown in the last line of the table shown in Fig.~\ref{fig:screenshot}. With Eqs.~\ref{eq:weight} and~\ref{eq:weighted} applied to the coupling precisions indicated therein (ignoring the coupling to the c quark and taking the same Higgs self coupling precision for CEPC as for FCC-ee, as instructed in Ref.~\cite{Breidenbach:2023nxd}), we find the average couplings summarised in Table~\ref{tab:weighted}. They vastly differ from the published values and would thus modify the conclusion of Ref.~\cite{Breidenbach:2023nxd} that C$^3$ is {\it the most environmentally friendly option when the precision-weighted total carbon footprint is accounted for}.

\begin{table}[h]
\renewcommand{\arraystretch}{1.25}
\centering
\caption{\label{tab:weighted} \small Values of the weighted averages for the Higgs factories considered in Ref.~\cite{Breidenbach:2023nxd}, in the original table and in our calculation from Eqs.~\ref{eq:weight} and~\ref{eq:weighted}.}
    \begin{tabular}{c|c|c|c|c|c}
    \hline 
     Collider &  CLIC$_{380}$ & ILC$_{250}$ & ILC$_{500}$/C$^3_{550}$ & FCC-ee & CEPC \\ \hline
     Ref~\cite{Breidenbach:2023nxd} & 0.94  & 0.86 & 0.45 & 0.59 & 0.49 \\ \hline
     This work & 0.91 & 0.90 & 1.21 & 1.21 & 0.81 \\ \hline
   \end{tabular}
\end{table}
\noindent After contacting one of the authors of Ref.~\cite{Breidenbach:2023nxd}, who shared with us a screenshot of the table used as an input to their code, and supposedly identical to that in Fig.~\ref{fig:screenshot}, we realised that the Higgs-self coupling (hhh) precisions had mistakenly been divided by a factor 100 in their derivation (e.g., 0.50\% instead of 50\% for HL-LHC).\footnote{To get the values reported in Ref.~\cite{Breidenbach:2023nxd}, it is also necessary to impose that the HZ$\gamma$ precision values be bounded from above by the HL-LHC value.} This bug is corrected in the following (and in the last row of Table~\ref{tab:weighted}).
\vskip 0.10cm
\item We then found out that the metric proposed in Eqs.~\ref{eq:weight} and~\ref{eq:weighted} is not mathematically sound (i.e., it is not a metric), in the sense it is not a monotonic function of the coupling precision: two very different values for the precision can give the same average weight. This feature is illustrated in Fig.~\ref{fig:monotonic} for ILC$_{500}$. 
\begin{figure}[h]
\centering
\includegraphics[width=0.95\textwidth]{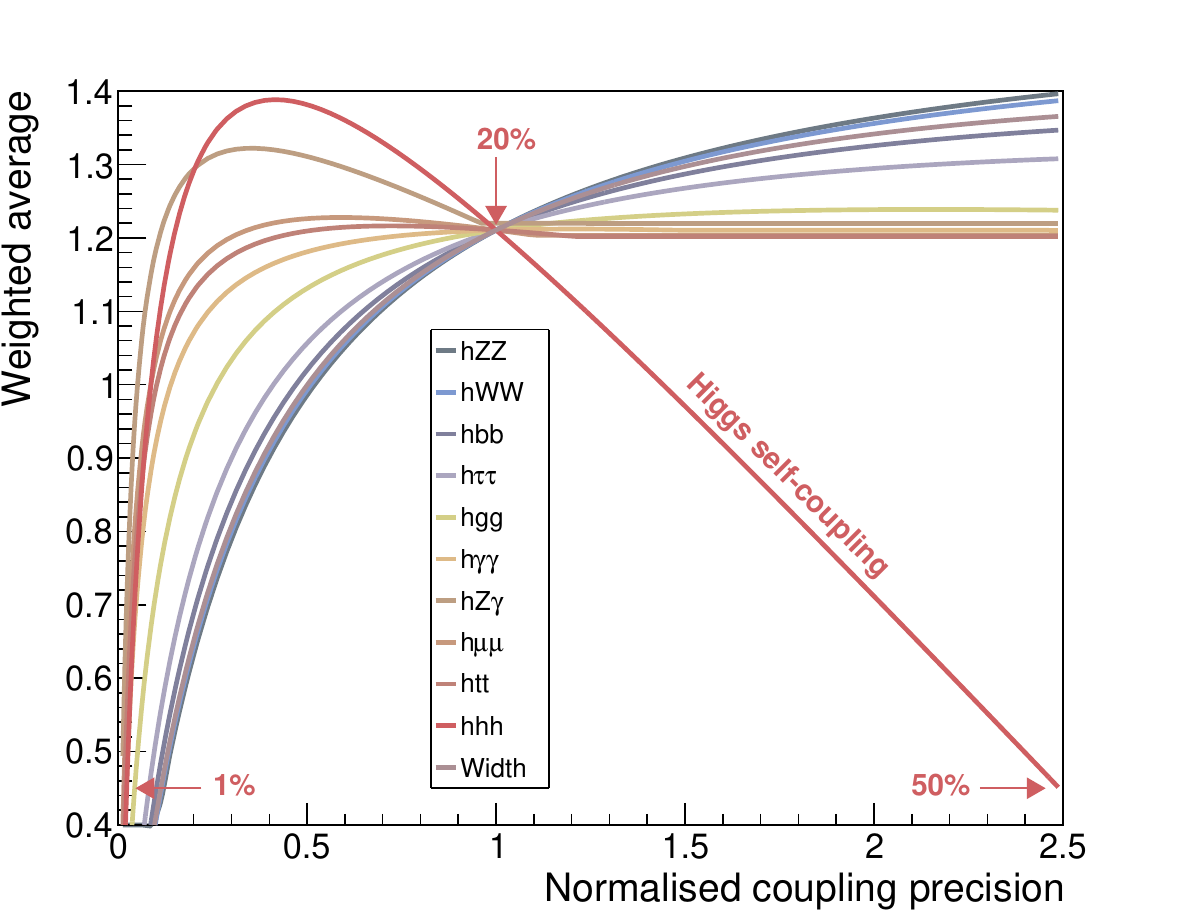}
\caption{\label{fig:monotonic} \small Evolution of the weighted average of ILC$_{500}$ when varying individually the coupling (or width) precisions from 0.025 to 2.5 times their default value indicated in Fig.~\ref{fig:screenshot}, all other precision values remaining the same as in Fig.~\ref{fig:screenshot}. The ``hhh'' (red-ish) curve shows the contribution to the weighted average from the Higgs self-coupling precision, given to be 50\% for HL-LHC and 20\% for ILC$_{500}$ in Fig.~\ref{fig:screenshot} (but mistakenly changed to 0.5\% and 0.2\% in Ref.~\cite{Breidenbach:2023nxd}): the weighted average degrades when the self-coupling precision improves from 50\% to 10\%, exhibits a maximum for a precision around 10\%, and becomes better than no measurement at all only for precisions of the order of 1\% or below, although it is supposed to be an estimator of the precision improvement with respect to HL-LHC.}
\end{figure}
For example, it can be seen that a precision of 20\% for the Higgs self-coupling (red-ish curve in Fig.~\ref{fig:monotonic}, normalized coupling precision equal to unity) at ILC$_{500}$ actually increases (by a factor close to 3) the weighted average with respect to no measurement at all (normalised coupling precision of 2.5), and that a precision below 1\% is required to start improving the weighted average. This is a serious flaw that invalidates the conclusions drawn in Ref.~\cite{Breidenbach:2023nxd} with this estimator.
\vskip 0.10cm
\item Another weakness of the proposed ``metric'' is that it is not able to cope with couplings beyond those accessible at HL-LHC (e.g., hc\=c, hs\=s, h$\rm e^+e^-$). Such couplings provide infinite knowledge improvement, and and definitely ought to be included in the weighted average.
\vskip 0.1cm
\item Mathematically-sound metrics based on the coupling precision improvement with respect to HL-LHC can easily be constructed. There are, in fact, many valid possibilities, rendering the choice somewhat arbitrary, as it may lead to tunable conclusions. Moreover, the physics meaning of such metrics is debatable, as the improvement with respect to HL-LHC of such or such coupling, even if it is indeed an important information, may not say much about the corresponding sensitivity to new physics. For example, having a measurement of the Higgs self-coupling with a precision of 20\%, even if it would improve the HL-LHC measurement by a factor 2.5, might bring very little additional constraint to many new physics theories, especially when a precision of the order of the per mil can be reached on other couplings. 
\vskip 0.1cm
\item In case the authors of Ref.~\cite{Breidenbach:2023nxd} wanted to fix their ``metric'' for the aforementioned issues, they should also fix the table of precision displayed in Fig.~\ref{fig:screenshot}, according to the following list of inaccuracies.
\vskip 0.15cm
\begin{enumerate} 
\item The most important drawback in this table is, according to Ref.~\cite{Dawson:2022zbb} from which it is extracted, that the various columns do not refer to the same $\kappa$ fit: the linear collider columns refer to a kappa-0 fit, while the circular collider columns refer to a kappa-3 fit. This difference is pointed out clearly in the original table caption of Ref.~\cite{Dawson:2022zbb}. As explained in Ref.~\cite{deBlas:2019rxi}, the kappa-0 fit (here updated with respect to Ref.~\cite{deBlas:2019rxi} to include HL-LHC) constrains the invisible and non-standard Higgs decay branching ratios to be exactly 0, while the kappa-3 fit let them free, including only measurements of these quantities in the fit, if available (unlike what the table caption in Fig.~\ref{fig:screenshot} says). Because the former fit is much more constrained than the latter, it mechanically returns significantly better coupling precision, thus giving an artificial head start to linear colliders in the comparison of Ref.~\cite{Breidenbach:2023nxd}. For example, the  hZZ coupling at FCC-ee (with 2 IPs) would have a precision of 0.14\% in the kappa-0 fit instead of the 0.17\% quoted. 
\vskip 0.15 cm
\item In the kappa-0 fit, the total decay width is 100\% correlated with the fitted $\kappa$'s. It must therefore not be included in the metric for linear colliders, as it double-counts features already carried by the couplings. 
\vskip 0.15cm
\item Since the beginning of the FCC Feasibility Study (2021), i.e., two years before the publication of Ref.~\cite{Breidenbach:2023nxd}, the baseline FCC layout has been modified to be compatible with 4 IPs, and the 4 IP scenario has become the baseline for FCC-ee. The table of Fig.~\ref{fig:screenshot} is therefore obsolete. As it needs to be updated, the latest integrated luminosities of FCC-ee~\cite{janot_2024_nfs96-89q08} could be used as well. For example, the hZZ coupling at FCC-ee (with 4 IPs) would have a precision on 0.094\% in the kappa-0 fit~\cite{deBlas:2019rxi} instead of the 0.17\% quoted.\footnote{In fact, Table 30 of Ref.~\cite{deBlas:2019rxi}, published in 2019, i.e., 4 years before Ref.~\cite{Breidenbach:2023nxd}, contains all the results of the kappa-0 fit for FCC-ee in combination with HL-LHC, with 2IPs and 4IPs, and should therefore have been used in a fair comparison.}
\vskip 0.15cm
\item Some of the entries in the table are internally inconsistent. For instance, all hZ$\gamma$ coupling precisions at CLIC, ILC, and FCC, supposedly combined with HL-LHC, are worse than the HL-LHC corresponding precision. In addition, the hZ$\gamma$ precision of CEPC should probably be propagated to FCC-ee, as no analysis has been carried out on this channel by the FCC teams. 
\vskip 0.15cm
\item It is (incorrectly) assumed that the Higgs self-coupling can be measured with CEPC with the same precision as with FCC-ee, in spite of the very different integrated luminosities at 240\,GeV and at the t\=t threshold.  
\vskip 0.15cm
\item Finally, the $\rm e^+e^-$  collider's hhh precisions are obtained by a global SMEFT fit. It would have been more consistent to use the SMEFT results for all couplings.\footnote{The results of a SMEFT fit (internally consistent for all colliders) were also already available in Ref.~\cite{deBlas:2022ofj} for use in Ref.~\cite{Breidenbach:2023nxd}. This aspect is fully addressed in Ref.~\cite{Blondel:2024mry}}
\end{enumerate}
\end{enumerate}

We conclude from the above that the metric used in Ref.~\cite{Breidenbach:2023nxd} for the comparison of ``carbon footprint per unit of physics output''
is flawed in many respects, and gives an extremely skewed, and in fact incorrect, representation of reality. 

A different metric that would solve most issues (but item 3.) listed above, could be to define the combined improvement with respect to HL-LHC as the quadratic combination of all individual improvements:
\begin{equation}
    W = 100 \times \left( \sum_i \frac{ \displaystyle \left(  \frac{\delta\kappa}{\kappa} \right)^2_{\rm HL-LHC, i} } { \left( \displaystyle \frac{\delta\kappa}{\kappa} \right)^2_{\rm HL-LHC+HF, i} } \right)^{-1}.
    \label{eq:newmetric}
\end{equation}

\begin{figure}[h]
\centering
\includegraphics[width=0.95\textwidth]{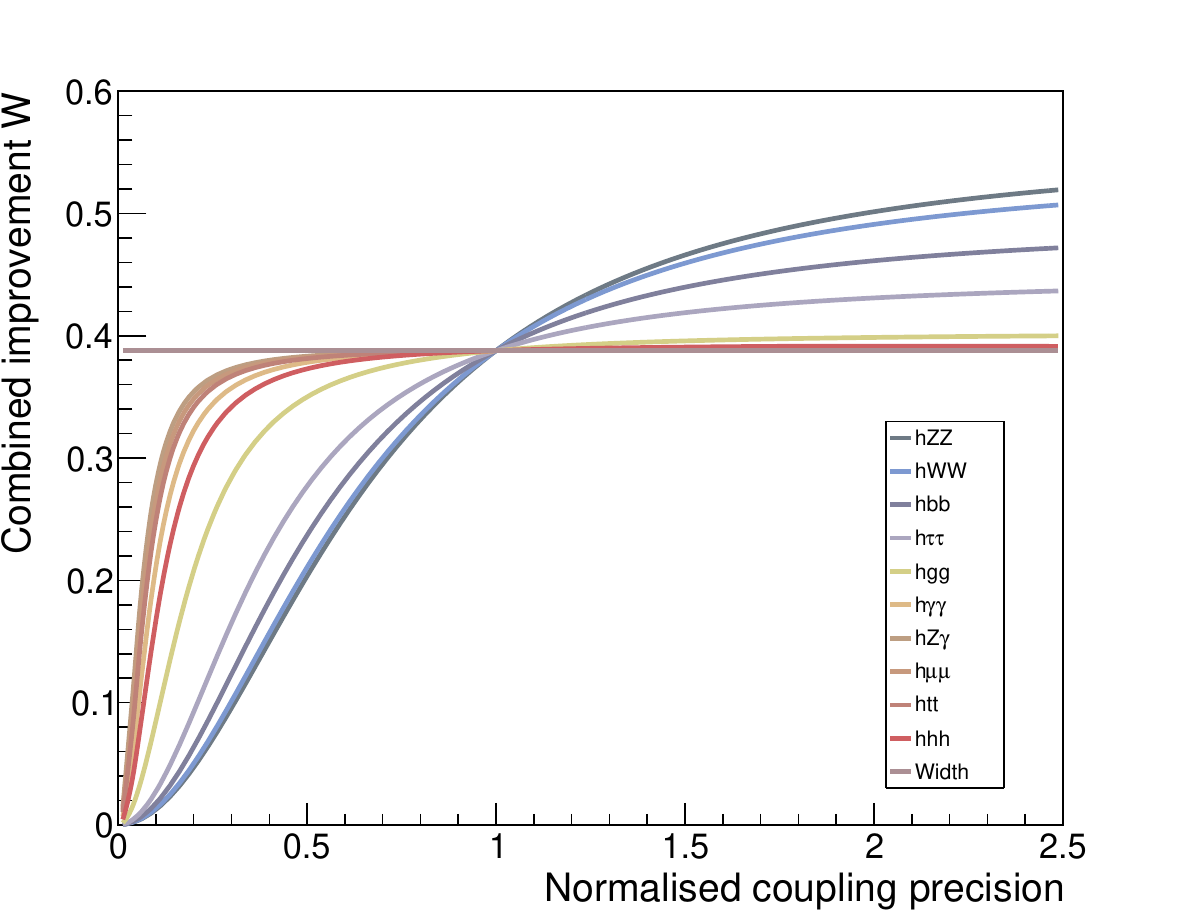}
\caption{\label{fig:newmetric} \small Evolution of the combined improvement $W$ with respect to HL-LHC defined in Eq.~\ref{eq:newmetric}, for ILC$_{500}$, when varying individually the coupling precisions from 0.025 to 2.5 times their default (corrected) value, all other coupling precision values staying constant. As explained in item 5(b) above, the total width has been removed from the definition of $W$, hence $W$ is independent of the corresponding precision.} 
\end{figure}

As displayed in Fig.~\ref{fig:newmetric}, this metric is indeed a monotonically decreasing function of the precisions, as reflected in the combined improvement value for each collider: $W_{\rm CEPC} = 0.17$, $W_{\rm FCCee} = 0.21$, $W_{\rm C^3_{550}} = 0.38$, $W_{\rm ILC_{500}} = 0.39$, $W_{\rm ILC_{250}} = 1.22$, and $W_{\rm CLIC_{380}} = 1.73$. (These values are obtained with a kappa-0 fit for all colliders.) It also gives less importance to the couplings that are not or little improved with respect to HL-LHC, as intended.\footnote{Incidentally, this metric does not depend on whether a specific coupling precision is expressed in per cent (e.g., 50\%) or in absolute (e.g., 0.5).} 

Sadly, this metric -- as any other metric -- would still be arbitrary, as it would allow infinite freedom in the choice for the collider weight. Indeed, any power of $W$ would have the same good mathematical properties, but would increase (for exponents larger than unity) or reduce (for exponents smaller than unity) the ratios between the collider weights, therefore leading to tunable -- thus not scientific -- conclusions when used as a multiplicative coefficient. 

\begin{figure}[h]
\centering
\includegraphics[width=1.0\textwidth]{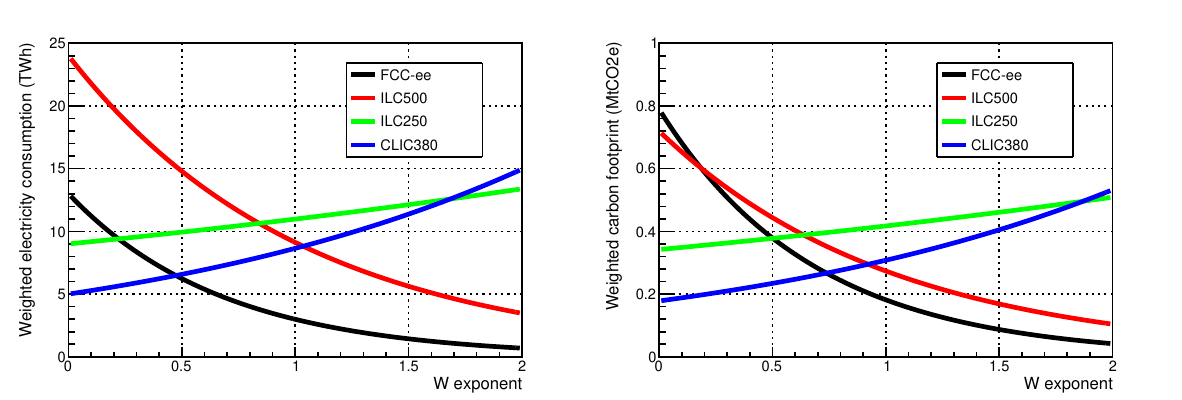}
\caption{\label{fig:whoisbest} \small Electricity consumption and carbon footprint ``per unit of physics output'' as a function of the metric exponent, for FCC-ee (black curve), ILC$_{500}$ (red curve), ILC$_{250}$ (green curve) and CLIC$_{380}$ (blue curve). A carbon intensity of 20\,kg CO$_2$e per MWh and the projected life cycle assessments of Refs.~\cite{footprintLC,johannes} are used for the respective carbon emissions at the time of construction and operation at CERN.}
\end{figure}

This undesired feature appears strikingly in Fig.~\ref{fig:whoisbest}, where the electricity consumption and carbon footprint ``per unit of physics output'' (i.e. multiplied by $W^x$) are shown as a function of $x$ between 0 and 2, for FCC-ee,  ILC$_{500}$, ILC$_{250}$ and CLIC$_{380}$: any ranking can be obtained as long as the {\it right} value of $x$ is chosen. 

We conclude that, irrespective of the specific inconsistencies and drawbacks of Ref.~\cite{Breidenbach:2023nxd}, the strategy consisting in using a metric {\it as a multiplicative coefficient} of the electricity consumption or the carbon footprint of an ensemble of colliders is fragile at best, and valueless when it comes to arguing in favour of or against such or such future Higgs factory.%

The above reasoning advocates strongly in favour of putting all colliders on an equal footing first, i.e., understanding with how much integrated luminosity they would all reach the same coupling precision, and therefore the same sensitivity to new physics coupled to the Higgs boson. This is the topic for another note~\cite{Blondel:2024mry}.

\small {\bf Note added:} We first contacted the authors of Ref.~\cite{Breidenbach:2023nxd} on 25 Nov 2024, to inform them that we could not reproduce their result. A first version of this note was communicated to them 10 days later. As their reply did not obviously indicate the possibility for an erratum to Ref.~\cite{Breidenbach:2023nxd}, and as it indeed did not happen after 10 more days, we thought that it was necessary to come out with this preprint, for informational purpose.   

\bibliography{References}
\end{document}